\documentclass[prl,aps,twocolumn,preprintnumbers,floats,amsmath,amssymb]{revtex4}

\usepackage{graphicx}% Include figure files
\usepackage{dcolumn}% Align table columns on decimal point
\usepackage{bm}% bold math
\usepackage{amssymb}% for unusual symbols
\usepackage[latin1]{inputenc}

\def \be{\begin{equation}}
\def \ee{\end{equation}}

\begin{document}

\title{A numerical finite size scaling approach to many-body localization}

\author{Geneviève Fleury and Xavier Waintal}
\affiliation{Nanoelectronics group, Service de Physique de l'Etat Condensé,
CEA Saclay F-91191 Gif-sur-Yvette Cedex, France\\}

\date{\today}

\begin{abstract}
We develop a numerical technique to study Anderson localization in interacting 
electronic systems. The ground state of the disordered system is calculated with
quantum Monte-Carlo simulations while the localization properties are extracted
from the ``Thouless conductance'' $g$, i.e. the curvature of the energy with respect
to an Aharonov-Bohm flux. We apply our method to polarized electrons in a two dimensional system of size $L$.
We recover the well known universal $\beta(g)=\rm{d}\log g/\rm{d}\log L$ one parameter scaling function without
interaction. Upon switching on the interaction, we find that $\beta(g)$ is unchanged while the
system flows toward the insulating limit. We conclude that polarized electrons in two dimensions
stay in an insulating state in the presence of weak to moderate electron-electron 
correlations.
\end{abstract}

\maketitle

Since the early days of Anderson localization \cite{anderson1958}, it is believed that 
in the thermodynamic limit, an arbitrary small disorder is enough to drive a two 
dimensional electron gas toward an insulator \cite{abrahams1979}. 
At the origin of this prediction is the scaling theory of localization \cite{abrahams1979}
which conjectured that the evolution of the conductance with the system 
size obeyed a simple one parameter scaling function. An important numerical effort has
since been devoted to establish the presence of this scaling \cite{pichard1981,kramer1993}
and calculate the scaling function. While this one electron localization picture is now
reasonably well understood, the corresponding many-body problem, where not only disorder
but also electron-electron interactions are considered, is yet unsolved.  An important litterature
has been devoted to the very strong \cite{efros1975} and weak disorder 
limit \cite{finkelshtein1983,castellani1984} but very little on the interplay between interaction and
localization itself~\cite{fleishman1980}.
It was generally assumed that electron-electron interactions did not modify drastically the one electron 
physics, so that the observation of a metallic state in two-dimensional Si MOFSETs in 1994~\cite{kravchenko1994}
came as an important surprise. It gave rise to a new interest in the subject~\cite{dobrosavljevic1997,waintal2000,caldara2000} and raised the question of the possibility that electron-electron interaction could stabilize a 
metallic phase. Recent progresses in the weak disorder limit seem to indicate that it could indeed be the 
case\cite{punnoose2005}.

 Numerical methods have proved to be very useful in putting the scaling theory of 
localization for non-interacting particles on very firm grounds. It is therefore
very tempting to try to develop similar approaches for the many-body problem
in spite of the intrinsic difficulties in dealing with correlations. Indeed, a number of
technical problems need to be overcome. (i) Obtaining the 
ground state of a  decently large number $N$ of correlated particles is already a challenging task.
(ii) In order to study Anderson localization (and not a mere trapping of the electrons which is
found for very strong disorder) one needs the localization length $\xi$ to be rather large 
yet smaller than the system size $L$ which must hence be rather large itself. 
(iii) One needs to calculate a physical
observable sensitive to localization which must hence be some sort of correlation function~\cite{basko2006}. Indeed, thermodynamic quantities such as the electronic density do not show localization in average.

\begin{figure}
\includegraphics[angle=270,keepaspectratio,width=8.5cm]{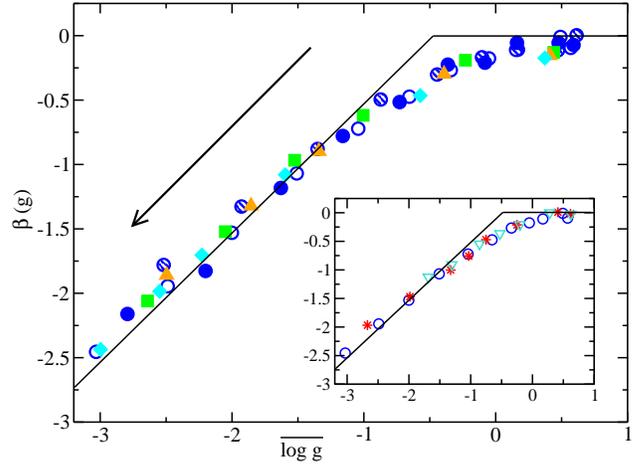}
\caption{\label{cbe_beta_de_g} (Color online). Scaling function $\beta(g)$ as a function of $\overline{\log g}$, 
in rectangular systems with $\nu=1/24$. The symbols correspond to different system sizes, $N=16$ 
particles in $16\times24$ sites (empty symbols), $N=25$ in $20\times30$ (full symbols), $N=36$ in $24\times36$ (striped symbols), and different strengths of the interaction, $r_s=0$ (circles), $r_s=2$ (squares), $r_s=4$ (up triangles), $r_s=6$ 
(diamonds), for various strengths of the disorder ($0 < r_w < 30$). The black lines are the 
expected asymptotic limits. Inset: idem for various filling factors, $\nu=1/24$ (circles), $\nu=1/54$ (stars) and $\nu=1/96$ (triangles down), with $N=16$ particles, at $r_s=0$. Upon increasing disorder or 
interaction, the system flows toward the insulating limit as indicated by the arrow.}
\end{figure}

\begin{figure}
\includegraphics[angle=270,keepaspectratio,width=8.5cm]{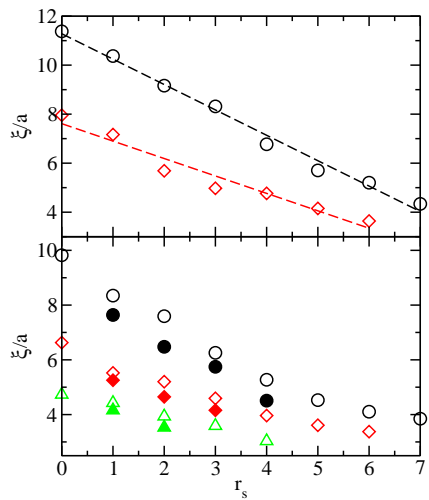}
\caption{\label{cbe_xsi_bilan} (Color online). $\xi/a$ as a function of $r_s$ for various disorders, $r_w=19.6$ (circles), $r_w=22$ (diamonds) and $r_w=24.5$ (triangles). Top: $N=25$ in $20\times30$ sites. Bottom: $N=16$ in $16\times24$. Empty (full) symbols correspond to $\Psi_{\rm Har}$ ($\Psi_{\rm liq}$).}
\end{figure}

In this letter, we propose a practical scheme to study numerically many-body localization.
We use zero temperature Green Function quantum Monte-Carlo (GFMC) technique to study the
ground state of the system. This technique takes full advantage of the fact that the non-interacting
ground state can be found exactly (by diagonalization of the one-body problem) so that upon
switching on the interaction the GFMC simulations are done with a very good starting point.
Our tool to measure the localization properties is the ``Thouless conductance'' of the system,
which can be related to the distribution of the winding numbers in the imaginary time path integral.
We apply our method to polarized (spinless) electrons in two dimensions for which both theory and experiments
agree that the system is insulating. The main point of scaling theory of localization is that $\beta(g)\equiv \rm{d}\log g/\rm{d}\log L$ which depends on disorder, interaction, density and size is in fact a function of $g$ only.
Our chief result is presented in Fig. \ref{cbe_beta_de_g} where we establish this scaling 
for interacting electrons.
We find that $\beta(g)$ is unaffected by the presence of the correlations due to Coulomb repulsion
in agreement with what is expected from the weak disorder limit~\cite{lee1985}.
Upon increasing the interaction strength, the system flows toward the insulating limit and the system 
localization length $\xi$ (shown in Fig.\ref{cbe_xsi_bilan}) decreases.

\begin{figure}
\includegraphics[angle=270,keepaspectratio,width=8.5cm]{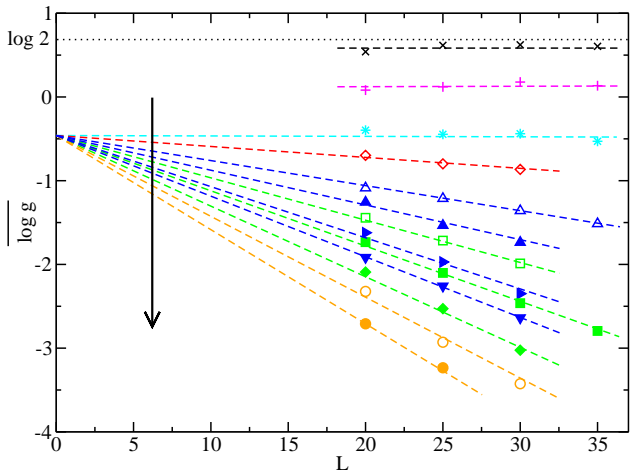}
\caption{\label{cbe_lng_vs_L} (Color online). $\overline{\log g}$ as a function of the system size $L$, in square systems with $\nu$=$1/25$. Disorder and interactions are increased from top to bottom as indicated by the arrow. The different curves are: $r_w$=0.005, $r_s$=0 ($\times$), $r_w$=10, $r_s$=0 ($+$), $r_w$=15, $r_s$=0 ($\ast$), $r_w$=17.5, $r_s$=0 ($\Diamond$), $r_w$=20, $r_s$=0 ($\triangle$), $r_w$=20, $r_s$=2 ($\blacktriangle$), $r_w$=20, $r_s$=4 ($\blacktriangleright$), $r_w$=20, $r_s$=6 ($\blacktriangledown$), $r_w$=22.5, $r_s$=0 ($\square$), $r_w$=22.5, $r_s$=2 ($\blacksquare$), $r_w$=22.5, $r_s$=4 ($\blacklozenge$), $r_w$=27.5, $r_s$=0 ($\circ$) and $r_w$=27.5, $r_s$=2 ($\bullet$). The collapse of the linear fits (dashed lines) at $L=0$ is the signature of one parameter scaling in the localized regime.}
\end{figure}

{\it Model and method.} 
We consider a system of $N$ spinless electrons in a rectangular $L_{x}\times L_{y}$ lattice with 
periodic boundary conditions. The Hamiltonian reads,
\be
\label{eq:model}
H=-t\sum_{\langle\vec r,\vec r'\rangle}c_{\vec r}^\dagger c_{\vec r'} + \sum_{\vec r}v_{\vec r}n_{\vec r}
+\frac{U}{2} \sum_{\vec r\ne\vec r'} V(\vec r-\vec r') n_{\vec r} n_{\vec r'}  + \lambda,
\ee
where $c_{\vec r}^\dagger$ et $c_{\vec r}$ are the usual creation and annihilation operators of 
one electron on site $\vec r$, the sum $\sum_{\langle\vec r,\vec r'\rangle}$ is restricted to nearest 
neighbors and $n_{\vec r}=c_{\vec r}^\dagger c_{\vec r}$ is the density operator. 
The disorder potential $v_{\vec r}$ is uniformly distributed inside $\left[-W/2,W/2\right]$. 
 $U$ is the effective strength of the two body interaction $V(\vec r)$. To reduce finite size 
effects, $V(\vec r)$ is obtained from the bare Coulomb interaction using the Ewald summation 
technique. The expressions for  $V(\vec r)$ has been given in \cite{waintal2006}. At small filling 
factor $\nu\equiv N/(L_x L_y)\ll 1$, we recover the continuum limit and we are left with two
dimensionless parameters, the usual $r_s=m^*e^2/(\hbar^2\epsilon\sqrt{\pi n})$ ($m^*$ effective mass,
$e$ electron charge, $\epsilon$ dielectric constant and $n$ electronic density) interaction 
parameter which for our model reads $r_s=U/(2 t \sqrt{\pi \nu})$ and a parameter 
$r_w=W/(t\sqrt{\nu})$ controlling the strength of the disorder. In the diffusive limit without 
interaction, the product of Fermi momentum $k_F$ by the mean free path $l$ is given 
by $k_F l = 192\pi/r_w^2$.

The GFMC method and our particular implementation has been given in \cite{waintal2006} to which we refer for details
and references. GFMC is a lattice version of the standard zero-temperature quantum Monte-Carlo 
methods (like diffusive quantum Monte-Carlo) that have enjoyed important success for both bosonic and
fermionic systems~\cite{foulkes2001}. Its principle is to project an initial variational guiding wave-function (GWF) $|\Psi_G\rangle$ onto the exact ground state $|\Psi_0\rangle$ by applying the 
projector operator $e^{-\beta H}$ in a stochastic way. Quantum Monte-Carlo
methods suffer from the so called sign problem when dealing with fermionic statistics. One
way out of the sign problem which has been quite successful is the fixed node 
approximation~\cite{foulkes2001} where upon projection onto $|\Psi_0\rangle$ the sign of the
wave-function is kept fixed. The method is variational and calculates the best wave-function 
compatible with the nodal structure of the GWF. Important effort is usually spent looking
for a GWF as close to the real ground state as possible. In the present case however, we can obtain
the ground state without interaction exactly by diagonalizing the corresponding one-body problem.
As we are interested in the evolution of the localization properties upon switching on the interaction,
we have an excellent starting point to begin with. The general form of our GWF is a Slater determinant 
multiplied by a Jastrow function,
\be
\label{eq:GWF}
\Psi_{G}(\vec r_1,\vec r_2 ...\vec r_N)=
{\rm Det\ }[\phi_i(\vec r_j)]\times \prod_{i<j} J(| \vec r_i - \vec r_j|).
\ee
The Jastrow part introduces some correlation and account for Coulomb repulsion. We use modified Yukawa 
functions \cite{stevens1973} : $J(r)=\exp[-\frac{aA(r_s)}{r}(1-e^{-B(r_s) r/a})]$ where $a=1/\sqrt{\pi\nu}$ is 
the average distance between electrons. $A(r_s)$ and $B(r_s)$ are variational parameters that we optimize 
while imposing the cusp condition $B=\sqrt{r_s/A}$ to reproduce the short distance behaviour. The nodal 
structure of the GWF depends only on the Slater determinant ${\rm Det\ }[\phi_i(\vec r_j)]$ which enforces the antisymmetry. The Slater determinant is constructed out of one-body orbitals $\phi_i$ that are obtained in 
two different ways leading respectively to $\Psi_{\rm liq}$ and $\Psi_{\rm Har}$ GWF. 
The orbitals of $\Psi_{\rm liq}$ are calculated by exact diagonalization of the one-body (disordered) problem
so that  $\Psi_{\rm liq}$  coincides whith the exact ground state without interaction ($r_s=0$).
The calculation of the orbitals of $\Psi_{\rm Har}$ proceeds in a similar way but we include iteratively 
the (Hartree) mean field potential due to the density $\langle n_{\vec r}\rangle$ of electrons 
in the one-body problem. The Hartree potential tends to screen the disorder leading to an increase of the GWF's localization length. We shall verify however that the GFMC results are not sensitive to the choice of GWF. 

{\it Measuring the localization properties.} The idea to use the sensibility $D$ of the system to a tilt in 
its boundary conditions as a criteria of localization was introduced very early by Edwards and 
Thouless~\cite{edwards1972}. Indeed, in a periodic system, the position of the boundary can be moved 
by a simple gauge transformation so that all sites are equivalent with respect to the boundary and a localized state is expected to have an exponentially small sensitivity to the boundary. More precisely, in presence of a 
small Aharonov-Bohm flux $\phi$, a current $I=-\partial E/\partial\phi$ 
flows in the system ($E$ is the total energy). When the flux is small, we have $I=-D\phi$ where 
$D=\partial^2 E/\partial\phi^2|_{\phi=0}$ is the curvature of the energy. This quantity $D$ is referred as 
the ``Thouless conductance'', the Drude weight, the conductivity stiffness or the superfluid stiffness depending 
on the context. For bosonic systems, $D$ is simply related to the superfluid fraction~\cite{pollock1987}. 
For fermions, it is related to the low frequency limit of the imaginary part of the conductivity~\cite{kohn1964}. 
For disordered system the product of $D$ with the density of states is proportional to the conductance of
the system~\cite{edwards1972,braun1997}.
In most cases, $D$ is positive and the system is diamagnetic but in some instances (one dimensional systems
with even number of spinless electrons~\cite{leggett1991}, or two dimensional systems with degenerate 
ground states~\cite{fye1991}) a paramagnetic ($D<0$) response can been found so that the widely used 
interpretation of $D$ as a conductance can sometimes be problematic. In anycase, it is a good measure of the 
localization properties of the system.

\begin{figure}
\includegraphics[angle=270,keepaspectratio,width=8.5cm]{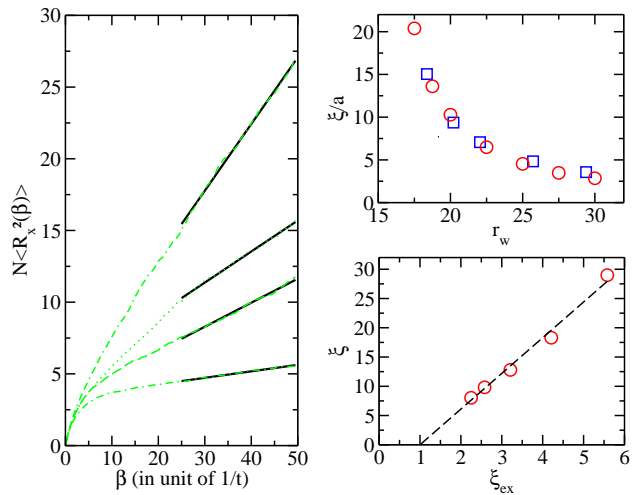}
\caption{\label{cbe_technique} (Color online). Left: example of $\langle R_{x}^{2}(\beta) \rangle$ for four different samples, $r_w=20$ and $r_s=4$.  The thick lines are the fits used to extract the values of $g$. Upper right: $\xi/a$ as a function of $r_w$, for $\nu=1/25$ (circles) and $\nu=1/54$ (squares) at $r_s=0$. Lower right: $\xi$ calculated with GFMC as a function of the exact result $\xi_{ex}$ at $r_s=0$ for $\nu=1/25$ and 
$20\le r_w\le 30$. The dashed line is a linear fit.}
\end{figure}

Following \cite{pollock1987}, we calculate the diffusive constant $g$ of the motion of the center 
of mass of the system in imaginary time along the $x$ direction, 
$g=\lim_{\beta\gg1} N \langle R_x^{2}(\beta) \rangle /(t\beta)$, where $\langle R_{x}^{2}(\beta) \rangle$ is the second moment of the center of mass along $x$. An example of the calculation of $g$ is shown in the left panel
of Fig.\ref{cbe_technique}. $g$ is simply related to $D$ as~\cite{pollock1987} $g= D L_x^2/(Nt)$ yet
$g$ is easier to access in the simulations. We note that in the fixed node approximation, $g$ is always positive
by construction. In particular, in the absence of disorder $r_w=0$ and interaction $r_s=0$,
we find $g=2$, i.e. the sum of the curvature of the individual one-body levels~\cite{kohn1964}. 

{\it Numerical results: square samples.} We now turn to the numerics and show that $g$ is an appropriate measure of
localization. In Fig. \ref{cbe_lng_vs_L} we plot $\overline{\log g}$ (averaged upon disorder)
as a function of system size $L$ for square samples at fixed filling factor. 
Without interaction, scaling theory of localization~\cite{abrahams1979} predicts that $\overline{\log g}$ is independent of $L$ for small disorder 
(Ohm's law) while for strong disorder, $g$ decreases exponentially with $L$ so that 
$\overline{\log g}=\log g_0-L/\xi$.
The existence of a universal scaling law $\beta(g)$ in this case means that $\log g_0$ is just a constant,
independent of the disorder strength. Indeed, the numerics are fully consistent with this picture: for $r_w\le 15$, 
$\overline{\log g}$ is roughly constant. Upon increasing  disorder further, $\overline{\log g}$
starts to decrease linearly with $L$ while all curves intercept at a single point at $L=0$. We find
 $\log g_0 =-0.47\pm0.08$. Further check of the method can be done by comparing the localization
length $\xi$ with results $\xi_{ex}$ of an exact diagonalization of the one-body Hamiltonian~\cite{explainXsiInf}. 
The result is shown in the lower right panel of Fig. \ref{cbe_technique}. We find both methods in
good agreement as $\xi\propto(\xi_{ex}-1)$~\cite{explainPropXsi}. In the upper right panel of Fig. \ref{cbe_technique}, we plot
$\xi/a$ for various values of $W$ and $\nu$ and verify that it is a function of $r_w=W/(t\sqrt{\nu})$.

We are now ready to switch on the interaction in our system. We find (Fig. \ref{cbe_lng_vs_L}, full symbols) that upon increasing
$r_s$, the localization length decreases so that the system becomes more insulating.
More importantly, all the curves still intercept at the same single point at $L=0$ indicating that the universal
scaling function $\beta(g)$ is unaffected by the electron-electron correlations.
To confirm this important point, we have performed simulations with our two different wave-functions 
$\Psi_{\rm liq}$ and $\Psi_{\rm Har}$ and find that the resulting $\xi$ agree very well as shown in the lowest panel of Fig. \ref{cbe_xsi_bilan}. This is a good test of the robustness of the method: even with $\Psi_{\rm Har}$
who tends to be {\it less} localized than $\Psi_{\rm liq}$, the FN-GFMC algorithm succeeds to find the (more localized) ground state. 

{\it Numerical results: rectangular samples.} The scaling function  $\beta(g)=\mathrm{d}\log g/\mathrm{d}\log L$ can in principle be extracted from Fig. \ref{cbe_lng_vs_L} by finite differences. More precise results are obtained
using rectangular ($L_y=1.5L_x$) samples. We now calculate the diffusion constants $g_x$ and $g_y$ along the two different directions and compute $\beta(g)=\overline{\log(g_y/g_x)}/\log(L_y/L_x)$ as a function of $\overline{\log g}=\overline{\log(g_x g_y)}/2$. This scheme allows
us to obtain the full scaling curve for a single system by varying the disorder parameter $r_w$.
The result is shown in Fig. \ref{cbe_beta_de_g}, for various values of $N$, $r_w$ and $r_s$ while different values
of $\nu$ are shown in the inset. All data collapse on one single curve. We emphasize that no operation is needed
to obtain this collapse, Fig. \ref{cbe_beta_de_g} shows raw data. The asymptotic curves are simple straight 
lines of slope one and zero which intercept at $\log g_0$ 
(which has been extracted from Fig.~\ref{cbe_lng_vs_L}).
Small deviations from scaling is observed for the
smallest size $N=16$ at $r_s\ge 6$. Much larger deviations were found for $N=9$ particles (not
shown). For $N=25$ and higher, the collapse was perfect up to our statistical accuracy.
The corresponding localization lengths are plotted in Fig. \ref{cbe_xsi_bilan}. They decrease nearly linearly with $r_s$, up to $r_s =7$. Although it is very difficult to reach higher values of $r_s$, it is very likely that the localization length stays below the non interacting one which rules out the possibility of a metallic behaviour.

{\it Conclusion.} We have proposed a practical method to study Anderson localization in presence of
many-body correlations. For polarized two-dimensional electrons we find that the universal scaling
function $\beta (g)$ is unaffected by the interactions. Yet, upon increasing the interaction strength,
the system flows toward the insulating limit. This picture is in agreement with what is expected in
the weak disorder and weak interaction limit~\cite{lee1985}. A natural extension of this work would be 
the study of non polarized electrons where the existence of an intrinsic metal-insulator transition remains a 
controversial issue. We note that in Fig. \ref{cbe_xsi_bilan}, the localization length $\xi$ extrapolates
to zero at $r_s\approx 10$ for the two studied values of disorder. This could be the signature of a transition
toward some sort of disordered Wigner crystal. Remarkably, $r_s\approx 10$ also corresponds to the density
at which the metal-insulator transition was observed for all but the cleanest samples~\cite{yoon1999}.

{\it Acknowledgment.} We thank G. Montambaux, J-L Pichard, F. Portier, P. Roche and K. Kazymyrenko 
for interesting discussions. 

\newcommand{{{\PRB}}}{{{Phys. Rev. B}}}\newcommand{{{\PRA}}}{{{Phys. Rev. A}}}\newcommand{{{\PRL}}}{{{Phys. Rev. Lett}}}\newcommand{{{\NPB}}}{{{Nucl. Phys.}}}\newcommand{{{\RMP}}}{{{Rev. Mod. Phys.}}}\newcommand{{{\ADV}}}{{{Adv. Phys.}}}\newcommand{{{\EPJB}}}{{{Eur. Phys. J. B}}}\newcommand{{{\EPL}}}{{{Eur. Phys. Lett.}}}


\begin{thebibliography}{79}
\bibitem{explainXsiInf}\bibinfo{note}{The infinite system localization length $\xi_{ex}$ is calculated by collapsing our 
 $\xi_{ex}(L)$ data on a single curve, according to the scaling law $\xi_{ex}(L)/L=f(\xi_{ex}/L)$.
 The one-body localization lengths $\xi_{ex}(L)$ are obtained from the square root of the participation 
 ratio of the one-body orbitals at the Fermi level: 
 $\xi_{ex}(L)=\overline{ \sqrt{1/\sum_{\vec r}|\phi_N(\vec r)|^4} }$}.
\bibitem{explainPropXsi}\bibinfo{note}{The minus one offset to $\xi_{ex}$ comes from the fact that for strong 
 disorders $\xi\to0$ while $\xi_{ex}\to1$}.
\end{thebibliography}
\end{document}